\documentclass[12pt,english,preprint,superscriptaddress]{revtex4-1}
\usepackage[T1]{fontenc}
\usepackage[latin9]{inputenc}
\usepackage{amsmath}
\usepackage{graphicx}
\usepackage{amssymb}

\makeatletter
\@ifundefined{textcolor}{}
{%
 \definecolor{BLACK}{gray}{0}
 \definecolor{WHITE}{gray}{1}
 \definecolor{RED}{rgb}{1,0,0}
 \definecolor{GREEN}{rgb}{0,1,0}
 \definecolor{BLUE}{rgb}{0,0,1}
 \definecolor{CYAN}{cmyk}{1,0,0,0}
 \definecolor{MAGENTA}{cmyk}{0,1,0,0}
 \definecolor{YELLOW}{cmyk}{0,0,1,0}
 }

\makeatother

\usepackage{babel}

\begin{document}

\title {Interaction Driven Interband Tunneling of Bosons in the Triple Well}

\author{Lushuai Cao}
\email{lcao@physnet.uni-hamburg.de}
\affiliation{Zentrum f\"{u}r Optische Quantentechnologien, Luruper Chaussee 149, D-22761 Hamburg, Germany}
\author{Ioannis Brouzos}
\email{ibrouzos@physnet.uni-hamburg.de}
\affiliation{Zentrum f\"{u}r Optische Quantentechnologien, Luruper Chaussee 149, D-22761 Hamburg, Germany}
\author{Sascha Z\"{o}llner}
\email{zoellner@nbi.dk}
\affiliation{Niels Bohr International Academy, Niels Bohr Institute, Blegdamsv$\epsilon$j 17, DK-2100 Copenhagen, Denmark}
\author{Peter Schmelcher}
\email{pschmelc@physnet.uni-hamburg.de}
\affiliation{Zentrum f\"{u}r Optische Quantentechnologien, Luruper Chaussee 149, D-22761 Hamburg, Germany}

\begin{abstract}
 We study the tunneling of a small ensemble of strongly repulsive bosons in a one-dimensional triple-well potential. The usual treatment within the single-band approximation suggests suppression of tunneling in the strong interaction regime. However, we show that several windows of enhanced tunneling are opened in this regime. This enhanced tunneling results from higher band contributions, and has the character of interband tunneling. It can give rise to various tunneling processes, such as single-boson tunneling and two-boson correlated tunneling of the ensemble of bosons, and is robust against deformations of the triple well potential. We introduce a basis of generalized number states including all contributing bands to explain the interband tunneling, and demonstrate various processes of interband tunneling and its robustness by numerically exact calculation.
\end{abstract}

\maketitle

\section{Introduction}
Ultracold atoms in optical lattices represent a rapidly-growing field of research \cite{review1,review2,review3}, with applications such as quantum simulators \cite{simulator} of strongly correlated systems, quantum phase transitions, such as the superfluid-Mott insulator transition \cite{Mott1,Mott2,Mott3}, or quantum computing and quantum information processing \cite{quan-info,quan-comp}. All these applications are rooted in the possibility of an extensive control of the underlying setup: the geometry and strength of trapping potentials can be tuned by the laser beams forming the lattice potential \cite{lattice-param0,lattice-param1,lattice-param2}, while the interaction strength can be changed via Feshbach resonances \cite{Feshbach-optical,Feshbach,Feshbach-magnetical}. Particularly in quasi one-dimensional systems, where the transversal degrees of freedom are energetically frozen by a tight transversal confinement, the effective interaction strength can be tuned by the confinement strength, leading to so-called confinement induced resonances \cite{cir,CIR2,CIR3,CIR4,CIR5}. As a hallmark example, this has led to the realization of the Tonks-Girardeau gas, a highly correlated one-dimensional bosonic state where the strong repulsive interaction induces fermionic properties into bosons \cite{TG-gas1,TG-gas2,TG-exp1,TG-exp2}.

The double well is the simplest case of a multiwell potential, and a plethora of phenomena have been observed and analyzed in great detail for this system. Loaded with a Bose-Einstein Condensate (BEC), the double well manifests itself as a bosonic analogue of the  superconducting Josephson junction \cite{Jos-exp,exact}. In the low-interaction regime, the coherent phase coupling of the BEC in neighboring wells dominates the tunneling properties, and gives rise to population transfer such as Rabi oscillations and $\pi$-mode oscillations \cite{pi-mode}, while in the somewhat stronger-interaction regime, the nonlinearity introduced by the interaction dominates, and the BEC can become self-trapped \cite{MQST,Self-trapping-exp}. On the other hand, for few-body systems, the microscopic counterpart of BECs, the tunneling between two neighboring wells also takes place in a correlated manner, exhibiting pair tunneling in the low- to intermediate-interaction regime \cite{pair-exp,pair-tunneling-tho}, and turning to self-trapping for stronger interactions \cite{pair-tunneling-tho}, which indicates the intrinsic correlation between micro- and macro-ensembles of ultracold atoms. The generalization from double well to multiwell systems can provide a bottom-up understanding of mechanisms operating in the infinite optical lattice. For instance, the study of the ground state properties by increasing the interaction strength in the double well and multiwells with the well number up to seven shows a common behavior of vanishing inter-well correlations \cite{double1,multi,correlation}, which is related to the superfluid-Mott insulator transition in the infinite optical lattice.

 A straightforward natural extension of the double well is the one-dimensional triple well. In the triple well, ultracold atoms also show correlated tunneling and self-trapping in the lower- and stronger interaction regime, respectively \cite{triple1}, and present at the same time novel properties such as stationary tunneling, in which the loss of coherence between bosons leads to a steady state \cite{stationary}. Reference \cite{triple2} provides an extensive study of the long-time evolution of the dynamical properties of micro- and macro-ensembles of bosons in the triple well. All these studies may generalize the phenomena in the double well to the optical lattices, and bridge the gap between the micro- and macro-systems, as we can see the triple well as a protype of optical lattices. Moreover, similarly to the analogue of the double well to the Josephson junction, the triple well is the minimal system which can model the source-gate-drain junction, and draws lots of attention from the perspective of atomtronics. A variety of proposals to achieve controllable atom transport based on the triple well have been presented \cite{atomtronics,atomtronics2,transistor}.

A widely used approximation in the study of cold atoms is the single-band approximation, which ignores the higher band contribution. This single-band approximation, working successfully in the low interaction regime, finds difficulties in the strong interaction regime, where the strong correlation between bosons suppresses the coherence, and consequently leads to phenomena such as the fragmentation of bosonic ensembles \cite{fragmentation}, fragmented pair tunneling and eventually breaking up of the tunneling pairs in the Tonks-Girardeau regime \cite{pair-tunneling-tho,fragmented-pair}. Efforts have been done to extend the studies to the strong-interaction regime, by explicitly taking into account the higher band contributions. It has been shown that higher band effects can give rise to effective many-body interactions \cite{high-interaction,high-interaction-exp}, and can modify the Bloch oscillation \cite{high-bloch} as well as Mott transition \cite{high-mott,multi,zoo,populationBH}.

In this work, we study the dynamical properties of a few-boson system in a one-dimensional triple-well with the numerically exact Multi-Configuration Time-Dependent Hartree Method (MCTDH) \cite{MCTDH1,MCTDH2,MCTDH3}. In the strong interaction regime, where it is common sense that tunneling in the multiwell is suppressed, we observe that multiple windows of enhanced tunneling are opened on top of the suppressed-tunneling background. These tunneling revivals result from the resonant coupling of the initial state to particular higher band states, and this enhanced tunneling is interband tunneling. Various tunneling patterns such as single-boson and correlated tunneling can be realized in corresponding windows of enhanced tunneling, which could be used to transport cold atoms in a controllable way.

The present work is organized as follows: in Sec. II, our setup and the MCTDH method are introduced. In Sec. III we discuss the construction of generalized number states which include interactions and are ideally suited for the analysis of the interband tunneling. In Sec. IV we present our results and their discussion and interpretation, and demonstrate in particular the interband process. Sec. V contains the summary and conclusions.

\section{Setup and Computational method}
\subsection{Setup}
In this work, we explore the correlated quantum dynamics of a few-boson system confined to a one-dimensional triple well. The Hamiltonian takes on the appearance:
\begin{equation}{
H=\sum_{j=1}^{N}-\frac{\hbar^{2}}{2M}\partial_{x_{j}}^{2}+\sum_{j=1}^{N}V_{tr}(x_{j})+\frac{1}{2}\sum_{j\neq k}g_{1D}\delta(x_{j}-x_{k})
}.
\end{equation}
Here $V_{tr}(x)=V_{0}sin^{2}(\kappa x)$ models the $1D$ triple-well confinement, where for convenience we apply hard-wall boundary conditions at $x=\pm3\pi/2\kappa$ ($\kappa$ is the wave vector of the laser beams forming the optical lattice) so as to confine the bosons to three adjacent wells, as schematically shown in figure 1(a). Experimentally the triple well potential can be realized, $e.g.$, by a bichromatic optical lattice, formed by laser beams of different frequency, or a harmonic trap superimposing an optical lattice. The shape of the triple well could correspondingly deviate from the ideal sine-square form which we are using here. However our results and findings are to a large extent independent of the exact form the triple well potential and the boundary conditions, and similar effects would be detected for various experimentally realizable setups.

The last term of $H$ represents the 1D contact-interaction potential. Under cylindrical harmonic confinement, the effective 1D coupling becomes $g_{1D}=\frac{2\hbar^{2}a_{0}}{Ma_{\bot}^{2}}(1-\frac{|\zeta(1/2)|a_{0}}{\sqrt{2}a_{\bot}})^{-1}$ \cite{cir}, where $a_{0}$ is the free-space s-wave scattering length, and $a_{\bot}=(\hbar/M\omega_{\bot})^{1/2}$ denotes the transverse-confinement length in terms of the frequency of the transverse harmonic confinement, $\omega_{\bot}$. It is clear from the expression for $g_{1D}$ that the interaction strength can be tuned by either changing the scattering length $a_{0}$, or the transverse confinement frequency $\omega_{\bot}$, appealing to field-induced Feshbach resonances \cite{Feshbach-optical,Feshbach,Feshbach-magnetical}, or confinement induced resonance \cite{CIR2,CIR3}, respectively.

 \begin{figure}
\includegraphics[width=16cm]{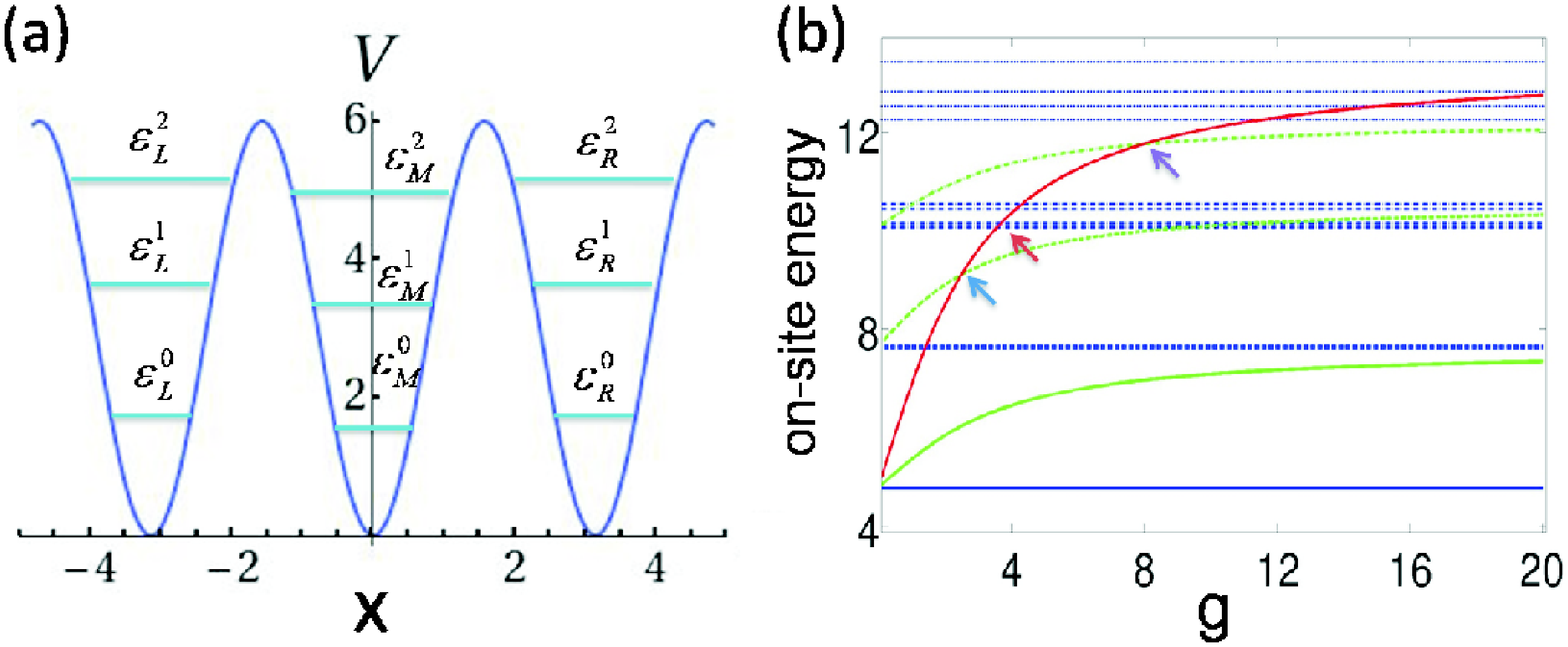}
\caption{(a) Sketch of the triple well potential as well as the first three Wannier levels in each well. (b) Response of the on-site energy of different number states to the interaction strength. The red line gives the response of the triple modes, and the blue and green lines provide the response of the single modes and pair modes, respectively. number states belonging to different bands are denoted by different line types: solid - ground band, dashed - first excited band,  dash-dotted - second excited band, dotted - third excited band. Arrows indicate the resonance points of the triple mode with various modes. } \label{figure 1}
\end{figure}

In the following we rescale (1) in units of the recoil energy $E_{R}=\hbar^{2}\kappa^{2}/2M$, while the space and time are given in the unit of $\kappa^{-1}$ and $\hbar/E_{R}$, respectively. This amounts to setting $\hbar=M=\kappa=1$. The coupling of the interaction potential now reads $g=2g_{1D}M/\hbar^{2}\kappa$. The triple well potential turns to $V_{tr}=V_{0}sin^{2}x$, with hard wall boundary conditions at $x=\pm3\pi/2$. In this work, $V_{0}$ is large enough such that each well confines three localized single-particle Wannier states: the ground, first excited and second excited Wannier states. As a result of the hard wall boundary conditions, the on-site energy of the Wannier states in the left and right well $\varepsilon^{i}_{L/R}$ is slightly larger than that of the Wannier states of the same energetic level in the middle well $\varepsilon^{i}_{M}$ ($i\in [0,2]$), as shown in figure 1(a).

\subsection{Computational Approach: The Multi-Configuration Time-Dependent Hartree Method}
We apply the numerically exact Multi-Configuration Time-Dependent Hartree $(\textbf{MCTDH})$ method \cite{MCTDH1,MCTDH2,MCTDH3}, a wave packet dynamical approach to the \emph{ab}-initio solution of multi-dimensional time-dependent Schr\"{o}dinger problems. In MCTDH the many-body wave function is expanded in terms of Hartree products of single particle functions of a corresponding basis:
\begin{equation}{
\Psi(x_{1},...,x_{N};t)=\sum_{j_{1},...,j_{N}}A_{j_{1},...,j_{N}}(t)\varphi_{j_{1}}(x_{1},t)...\varphi_{j_{N}}(x_{N},t)},
\end{equation}
where ${\varphi_{j_{i}}(x_{i})}$ are the single particle functions for the degree of freedom $x_{i}$. Substituting this Ansatz into the Schr\"{o}dinger equation, $i\hbar\partial_{t}\Psi=H\Psi$,  leads to a set of differential equations for  $A_{j_{1},...,j_{N}}(t)$ and $\varphi_{j_{i}}(x_{i},t)$. Integrating these differential equations, we obtain the time evolutions for arbitrary initial conditions $\Psi(x_{1},...,x_{N};0)$. MCTDH can also be used to calculate stationary states of given system by relaxation of a proper wave function, \textit{i.e.} imaginary time propagation. The advantage of MCTDH compared to exact diagonalization is that it optimizes the single particle function $\varphi_{j_{i}}(x_{i},t)$ at each time step, which leads to a reduction of the total number of necessary single particle functions in order to achieve convergence. When dealing with bosonic system, MCTDH can be simply modified to fulfill the permutation symmetry preserved by bosons. This can be done by symmetrizing the coefficients $A_{J}$, and using a single set of $\{\varphi_{j}(x,t)\}$ for all the degrees of freedom.

We aim at the tunneling dynamics of initial states where all bosons are localized in the same well. Such an initial condition can be experimentally prepared  \emph{e.g.} by adiabatically applying an external potential which rearranges the energies of minimum of the triple well, and all bosons are then confined to the lowest well, for instance, a linear tilt potential can confine all the bosons to the left well, and a harmonic potential originating at the middle well will localize all the bosons to the middle well; consequently this external potential is removed instantaneously. To prepare the initial state for our numerical calculations, we apply artificial hard wall boundary conditions on each side of the corresponding well, and let the bosonic wave function relax to the ground state inside the well via imaginary time propagation in the framework of MCTDH. Subsequently we turn off the artificial hard walls, and explore the time evolution $\Psi(x_{1},...,x_{N};t)$ in the triple well potential via MCTDH.

\section{Generalized Number-State Representation and Spatio \\-Energetic Characteristics of the Triple Well}

 Here we describe our number-state representations including the contribution of different bands. This yields an illustrative picture which helps us to analyze and interpret our numerical results obtained by MCTDH, and also serves as an explanatory tool. Especially the interband processes can be interpreted in the framework of this representation. We consider an ensemble of $N$ repulsively interacting bosons confined in the triple well potential as introduced in Sec. II.  The triple-well potential is deep enough such that the Wannier functions of the bosons belonging to different wells have very small overlap for not too high energetic excitation. The wave function of the bosons can be expanded in a series of generalized number states which denote the distribution of all the bosons among the three individual wells:

 \begin{equation}{
|\Psi\rangle=\sum_{\textbf{\emph{N}},i}C^{i}_{\textbf{\emph{N}}}|N_{L},N_{M},N_{R}\rangle_{i}
}.
\end{equation}
The summation runs over all possible configurations $\textbf{\emph{N}}=(N_{L},N_{M},N_{R})$ and the energetic excitations denoted by $i$. $N_{L}$, $N_{M}$ and $N_{R}$ are the number of bosons localized in the left, middle and right well, respectively, with $N_{L}+N_{M}+N_{R}=N$. We define the spatial representation of the number states as:
 \begin{equation}{
\langle \textbf{x}|N_{L},N_{M},N_{R}\rangle_{i}=S\phi^{i}_{\textbf{\emph{N}}}(\textbf{x}-\textbf{r}_{\textbf{\emph{N}}})
},
\end{equation}
  where $S$ refers to the symmetrization operation among all the bosons, and $\textbf{x}=(x_{1},...,x_{N})$, $\textbf{r}_{\textbf{\emph{N}}}=r_{L}^{N_{L}}\otimes r_{M}^{N_{M}}\otimes r_{R}^{N_{R}}$ ($\{r_{L},r_{M},r_{R}\}$ label the minimum positions of the three wells). The wave function $\phi^{i}_{\textbf{\emph{N}}}$ satisfies:
\begin{equation}{
(\sum^{N}_{j=1}-\frac{\hbar^{2}}{2M}\partial^{2}_{x_{j}}+\sum^{N}_{j=1}W(x_{j})+V_{I}(\textbf{x}))\phi^{i}_{\textbf{\emph{N}}}(\textbf{x})
=e^{i}_{\textbf{\emph{N}}}\phi^{i}_{\textbf{\emph{N}}}(\textbf{x})},
\end{equation}
where $W(x)=V_{0}sin^{2}(x)$ with hard wall boundary conditions at $x=\pm\pi/2$, is the local confining potential, and $e^{i}_{\textbf{\emph{N}}}$ is the $\emph{i}th$ on-site energy, with $i$ labeling the level of energetic excitation. The interaction potential in (5) takes the form:
\begin{equation}{
V_{I}(\textbf{x})=\frac{g}{2}[\sum_{j_{1}\neq j_{2}\in[1,N_{L}]}\delta(x_{j_{1}}-x_{j_{2}})+\sum_{j_{1}\neq j_{2}\in[N_{L}+1,N_{L}+N_{M}]}\delta(x_{j_{1}}-x_{j_{2}})+\sum_{j_{1}\neq j_{2}\in[N-N_{R}+1,N]}\delta(x_{j_{1}}-x_{j_{2}})]}.
\end{equation}

 A few remarks are in order here. Equation (5) is the constituting equation for the three subsets of bosons with particle number $N_{L}$, $N_{M}$ and $N_{R}$. Within each subset all interactions are taken into account whereas no interactions are taken into account for bosons belonging to different subsets, which is because we divide the triple well for our definition of number states into three adjacent single wells with hard wall boundary conditions. That is why $V_{I}$ can be divided into three terms according to (6). The number states $\langle \textbf{x}|N_{L},N_{M},N_{R}\rangle_{i}$ (see (4)) can be obtained from the wave functions $\phi^{i}_{\textbf{\emph{N}}}(\textbf{x})$ of (5), which are located in the interval $[-\pi/2,\pi/2]$, by performing the appropriate shift of arguments contained in $S\phi^{i}_{\textbf{\emph{N}}}(\textbf{x}-\textbf{r}_{\textbf{\emph{N}}})$, which leads to $N_{L}$, $N_{M}$ and $N_{R}$ bosons localized in the left ($[-3\pi/2,-\pi/2]$), middle ($[-\pi/2,\pi/2]$) and right well ($[\pi/2,3\pi/2]$), respectively.

 Let us discuss the response of the energies $e^{i}_{\textbf{\emph{N}}}$ to the interaction strength. As the interaction strength rises from zero to infinity, the wave functions of the number states change from the form of non-interacting bosons to that of fermionized bosons \cite{TG-gas1,TG-gas2,double1}, in which bosons with strong repulsive interaction in one dimension
  avoid spatial overlap and acquire fermionic properties. Correspondingly, the on-site energy increases from the sum of different single-boson Wannier energies and saturates to the corresponding value for non-interacting fermions. The latter can be described as $e^{i}_{\textbf{\emph{N}}}=e^{i}_{0}+e_{f}(N_{L},N_{M},N_{R})$, where $e^{i}_{0}$ is the value for non-interacting bosons, which is independent of the distribution of the bosons among the three wells, and increases with $i$. $e_{f}(N_{L},N_{M},N_{R})$ is the increase of the on-site energy in the fermionization regime compared to the non-interacting regime, which is dependent on the distribution of $(N_{L},N_{M},N_{R})$, $i.e.$, the more the bosons are localized to the same well, the higher $e_{f}(N_{L},N_{M},N_{R})$ is.

   When the on-site energy difference of two number states becomes much smaller than the effective coupling between them, significant tunneling will take place between them. For instance, the on-site energy of number states with lower $i$, but higher degree of localization of the bosons can cross that of other number states at a particular interaction strength, resulting in an enhanced tunneling between the corresponding number states. As such tunneling directly results from the contributions of excited number states, from higher bands of the system, we refer to this enhanced tunneling as "interband" tunneling.

 Let us briefly provide a comparison of the number states introduced here and in other works focusing on higher band effects. In previous works, the wave function is expanded with respect to the number states of non-interacting bosons, and the interaction potential is treated as a perturbation to these states. This means that these number states are just products of non-interacting single-particle Wannier functions. The perturbative treatment of the interaction potential limits such an expansion to the low-interaction regime. In our case the generation of number states intrinsically treats the interaction and the confinement potential, $i.e.$, $V_{I}(\textbf{x})$ and $W(x)$ in (5), in an inseparable manner, so that it becomes valid in the complete interaction regime, and uncovers the interplay between the interaction potential and the confinement potential. For instance, the interband tunneling discussed here can be easily understood in the number-state basis developed here but not in the number-state basis of non-interacting bosons applied in previous works.

\section{interband Tunneling: Observation and Analysis of Mechanism}

Interband tunneling is a general phenomenon insensitive to the total number of bosons, given all the bosons are well localized in the individual wells, and the minimal system to unravel this phenomenon is that of three repulsively interacting bosons in the triple well. In the following we demonstrate this process at hand of the system of three bosons in a triple well.

Firstly we briefly discuss the number-state properties of the three-boson system, especially the properties of the on-site energies of the number states. The number states can be divided into three categories: the "single mode" $\{|1,1,1\rangle_{i}\}$, implying that all the bosons are localized in different wells, the "pair mode" $\{|2,1,0\rangle_{i},|2,0,1\rangle_{i},|1,2,0\rangle_{i},|0,2,1\rangle_{i},|1,0,2\rangle_{i},|0,1,2\rangle_{i}\}$, refering to that two bosons are localized in the same well, and the "triple mode" $\{|3,0,0\rangle_{i}$, $|0,3,0\rangle_{i}$, $|0,0,3\rangle_{i}\}$, which refers to all the bosons residing in the same well. The three categories of number states show a different response to the increase of the interaction strength, as schematically shown in figure 1(b). For the single modes, the on-site energy is barely affected by the interaction strength and remains constant in the complete interaction regime as the summation of the on-site energy of single-boson Wannier states in each well, $i.e.$, $e^{i}_{|1,1,1\rangle}=\varepsilon^{l}_{L}+\varepsilon^{m}_{M}+\varepsilon^{n}_{R}$, where the configuration $(l,m,n)$ depends on the degree of energetic excitation $i$. The on-site energy of the pair modes always coincides with that of the corresponding single modes in the non-interacting regime, and saturates to the next upper band of the single modes for very strong interaction. For the triple modes, the on-site energy saturates to the third upper band of the single modes, and crosses several energy levels of single and pair modes in the intermediate regime. The on-site energy spectrum of the number states is shown in figure 1(b), in which we can see the "band-like" structure of the number states. In the figure, the "energy gap" between "bands" is to a good approximation the on-site energy difference of Wannier states of adjacent energy levels, while the energy splitting within a "band" results from the on-site energy difference of the Wannier states in the outer and middle well, $i.e.$, $\varepsilon^{l}_{L/R}-\varepsilon^{l}_{M}$. When the on-site energy of the triple modes crosses that of the single and pair modes of different energy levels, the interband tunneling process will take place, also marked in figure 1(b) with corresponding arrows. It is worth mentioning here that in the single band approximation, only number states with $i=0$ are considered, and the interband tunneling cannot be explained within this approximation.

In the following we provide several examples of the interband tunneling process by numerically exact quantum dynamical studies.

\subsection{Single-boson tunneling}

In this section, we demonstrate the single-boson interband tunneling of three bosons initially localized in the same well of a triple well. Let's concentrate first on the case where all atoms are loaded in the left (or equivalently rightmost) well, $\emph{i.e.}$, the initial state is of the form $|3,0,0\rangle_{0}$. The single-band model predicts self-trapping of all the bosons in the initial well when the interaction strength is large enough. Figure 2 shows the oscillation of the population in each well for the tunneling of three bosons initially localized in the left well for different interaction strengths. As shown in figure 2(a), for the interaction strength $g=0.1$, all the bosons remain essentially trapped in the left well, and tunneling is strongly suppressed. This indicates that the bosons become self-trapped due to the repulsive interaction. In fact, the local Lieb-Liniger parameter $\gamma=g/N\rho(x)$ for the initial state can be estimated as $\gamma\approx 0.03\ll1$, signalling weakly interacting bosons, and an effective lowest-band model is applicable, with  the local interaction energy $U_{0}\sim5.6$ and a hopping parameter $J\sim0.01$. Then we obtain $U_{0}/J\sim560\gg1$, which confirms the self-trapping phenomenon as shown in figure 2(a).

In figure 2(b), however, for $g=3.26$, where the single-band model predicts even stronger self-trapping than that of $g=0.1$, we observe the enhanced tunneling. At this value of $g$, the interaction parameter is $\gamma\approx1.2$, implying that on-site correlations are indeed non-negligible, and the single-band model is inapplicable here. In figure 2(b), the population in the left well oscillates between three and two, and that of the middle and right well oscillates between zero and one approximately. This implies that a single boson tunnels to the middle and right well with the other two bosons remaining localized in the left well.

\begin{figure}
\includegraphics[width=13cm]{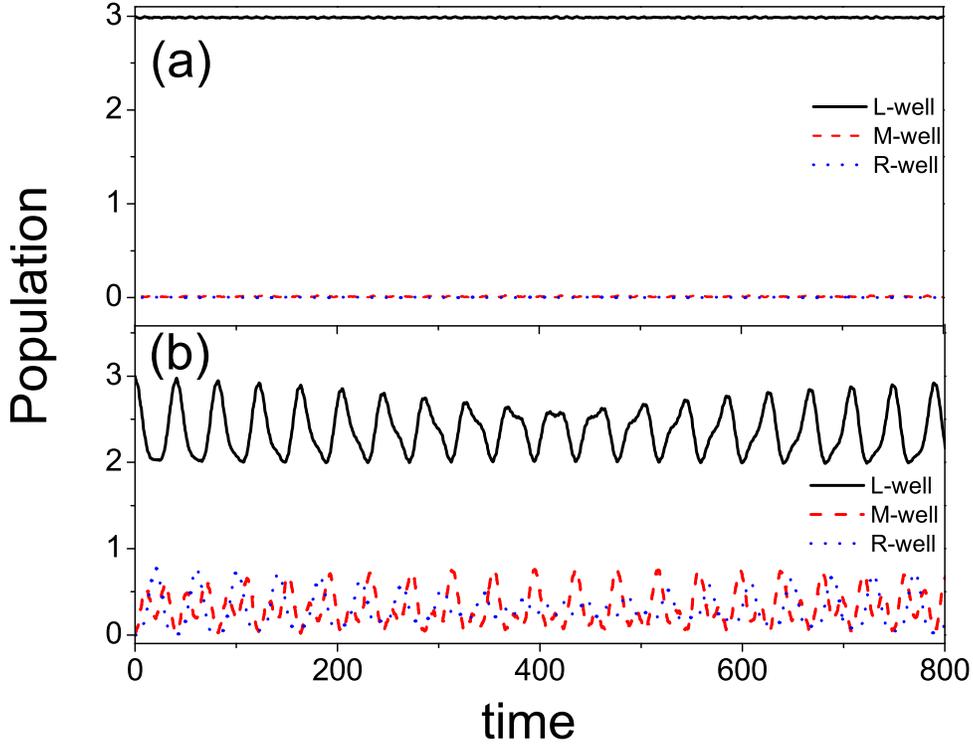}
\caption{The population oscillations of three bosons in the triple well system, which are initially localized in the left well.  The interaction strength is (a) $g=0.1$ and (b) $g=3.26$. Both figures demonstrate the emergence of enhanced tunneling windows on top of the suppressed-tunneling background, see discussion in text.} \label{figure 2}
\end{figure}

To clarify the mechanism underlying the enhanced tunneling in figure 2(b), we monitor the contributin from different number states, $|_{i}\langle N_{L},N_{M},N_{R}|\Psi(t)\rangle|^{2}$, as a function of time. Here the number states $|N_{L},N_{M},N_{R}\rangle_{i}$ are computed numerically utilizing improve relaxation within MCTDH, and then $|\Psi(t)\rangle$ is projected onto them at different times. The result is shown in Figure 3. Only the states $|3,0,0\rangle_{0}$, $\emph{i.e.}$ the initial state, and $|2,1,0\rangle_{1}$,$|2,0,1\rangle_{1}$ possess a significant contribution in the course of the tunneling process, indicating that the tunneling mainly takes place between these states. $|2,1,0\rangle_{1}$ and $|2,0,1\rangle_{1}$ refer to states where two bosons localize in the \emph{two-boson} ground state of the left well while one is in the first excited Wannier state of the middle and right well, respectively. In this way a boson hops to the first excited Wannier state of the middle and right well during the tunneling process, and the tunneling is indeed a single-boson tunneling.

 \begin{figure}
\includegraphics[width=12cm]{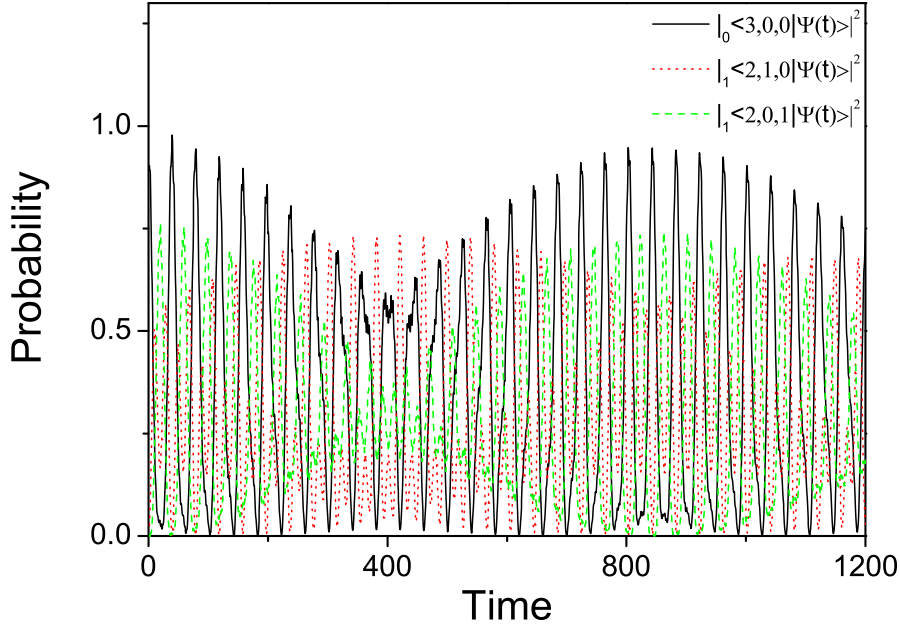}
\caption{ Probability of number states for the tunneling process shown in figure 2(b), for the states $|3,0,0\rangle_{0}$, $|2,1,0\rangle_{1}$ and $|2,0,1\rangle_{1}$, among which the tunneling mainly takes place. For this value of $g$, these three number states are in resonance, as marked by the blue arrow in figure 1(b).} \label{figure 3}
\end{figure}

 Besides the number states occupation, the tunneling process can also be demonstrated by the one-body density evolution. Figure 4(a) shows the one-body density at different time instants of the tunneling process. The density profiles can offer additional evidences of the excitation into different bands throughout the time evolution. At $t=0$ the density profile is roughly a Gaussian localized in the left well, and this corresponds to the initial state $|3,0,0\rangle_{0}$. As time evolves, the density profile populates the middle and right well, and instead of the typical single-particle ground-state profile of a Gaussian, the profile in the middle and right well shows a nodal structure, which is typical for a single boson in the first excited Wannier state, thereby verifying the contribution of the states $|2,1,0\rangle_{1}$ and $|2,0,1\rangle_{1}$ to which we referred above. The density profile appearing in the middle and right well is a spatial sign of the occupation of the first excited Wannier states in each of these wells, and based on this, we can give a schematic picture of this tunneling process as in figure 4(b), in which a single boson initially localized in the left well tunnels among all the three wells, and the other two bosons remain localized in the left well. When the boson tunnels to the middle and right well, it actually occupies the first-excited Wannier level in the corresponding well.

 \begin{figure}
\includegraphics[width=16cm]{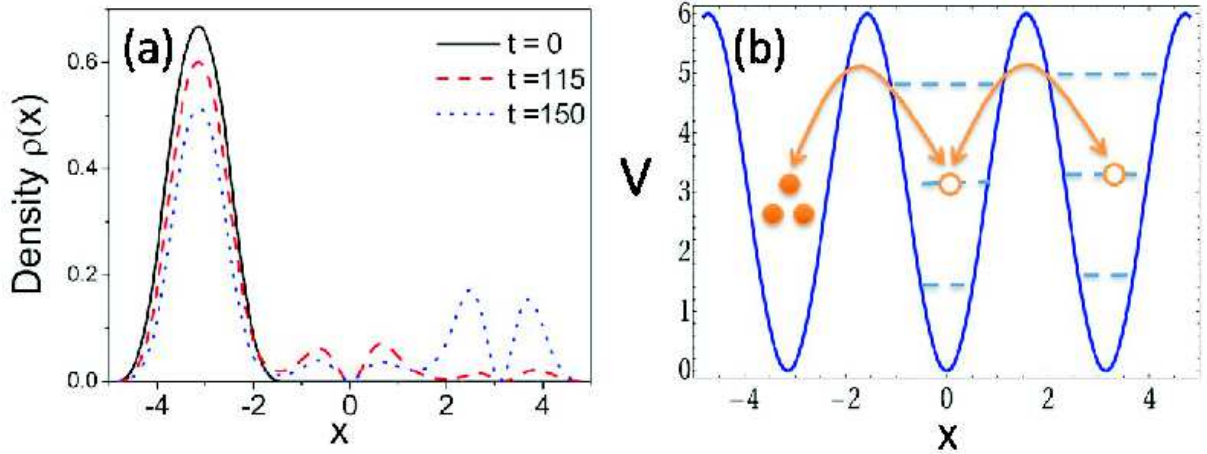}
\caption{(a) The one-body density at various time instants. Initially all bosons are localized in the left well (black line), then one boson tunnels to the middle well (red line), and to the right well (blue line). The nodal structure in the middle and right well demonstrates that the boson in the middle and right well are in the first excited Wannier level. (b) Visualization of the tunneling process: in the tunneling process a boson tunnels forth and back between the left, middle and right well, with the other two bosons remaining in the left well. The dashed lines in the middle and right well illustrate the Wannier energy levels as in figure 1(a). The parameters are the same as in figure 2(b) and figure 3.
} \label{figure 4}
\end{figure}

 Now we proceed to demonstrate an alternative single-boson tunneling, in which the tunneling boson can hop to even higher Wannier states at particular interaction strengths. As an example, we choose the initial state where all bosons are trapped in the middle well, $\emph{i.e.}$, $|0,3,0\rangle_{0}$, and the interaction strength $g=9.85$ corresponding to $\gamma\approx5.9$. We again calculate the time evolution of the population in each well and the probability of different number states, as shown in Figure 5. This time the population in the middle well oscillates approximately in the interval $[2,3]$, while the population in the left and right well oscillates approximately within $[0,0.5]$, as shown in figure 5(a). This indicates that a boson in the middle well tunnels to left and right well with the same probability. Figure 5(b) confirms this tunneling behavior with the probability evolution of the number states, in which the states $|0,3,0\rangle_{0}$, $|1,2,0\rangle_{3}$ and $|0,2,1\rangle_{3}$ share the major contribution and consequently tunneling mainly takes place among these number states. $|1,2,0\rangle_{3}$ and $|0,2,1\rangle_{3}$ refer to the states of two bosons in the middle well with the third one in the second excited Wannier state of the left and right well respectively, and this indicates that the bosonic tunneling to the left and right well hops to the second excited Wannier state.

\begin{figure}
\includegraphics[width=13cm]{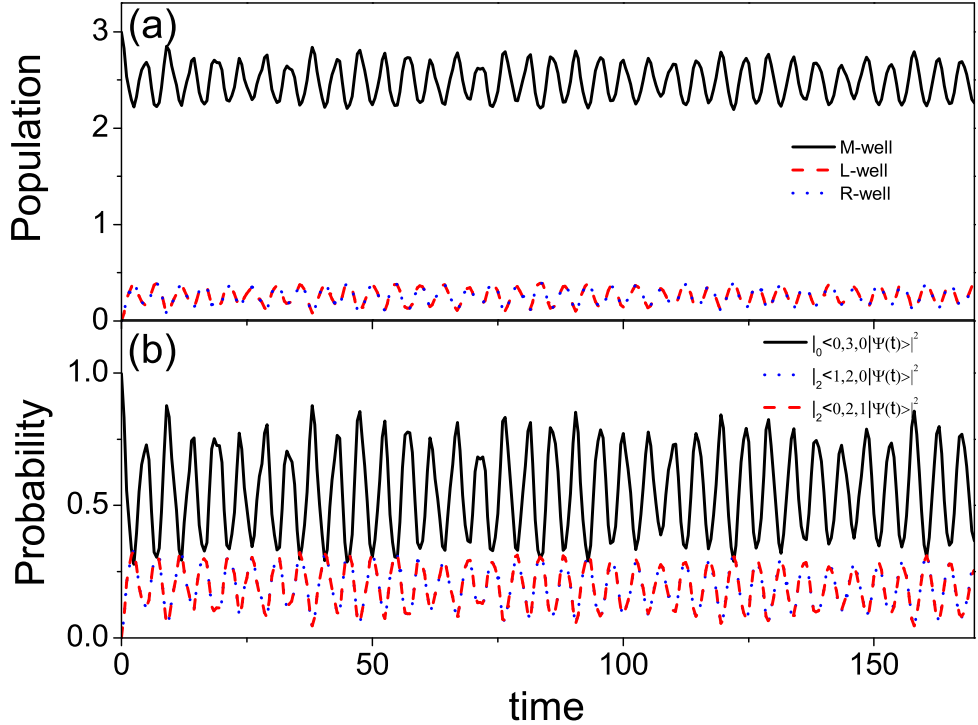}
\caption{(a) The population oscillation of three bosons in the triple well, for the interaction strength $g=9.85$. The three bosons are initially localized in the middle well. (b) Time-dependence of the occupation of $|0,3,0\rangle_{0}$, $|1,2,0\rangle_{3}$ and $|0,2,1\rangle_{3}$. As the system tunnels to the cat state $|1,2,0\rangle_{3}+|0,2,1\rangle_{3}$, one boson tunnels to the left and right well with the same probability, which gives rise to the coincidence of the population in the left and right well and that of the two number states $|1,2,0\rangle_{3}$,$|0,2,1\rangle_{3}$, as shown in (a) and (b), respectively.
} \label{figure 5}
\end{figure}

 To confirm and further analyze the above statement we show the corresponding one-body density for different times in figure 6(a). In the one-body density results, we see initially all bosons are trapped in the middle well. The deviation of this profile from a Gaussian indicates the onset of the fermionization process of the three strongly repulsively interacting bosons, which is similar to that of three strongly repulsively interacting bosons confined in the harmonic trap \cite{double1,harmonic1,harmonic2}. At later times, where a single boson tunnels to the left and right well, there are two nodes in these wells demonstrating the occupation of the second excited Wannier energy levels in both wells. We also illustrate this tunneling process in figure 6(b), and in this case the ensemble of bosons tunnels between the initial state $|0,3,0\rangle_{0}$, and the cat state $|1,2,0\rangle_{3}+|0,2,1\rangle_{3}$.

\begin{figure}
\includegraphics[width=16cm]{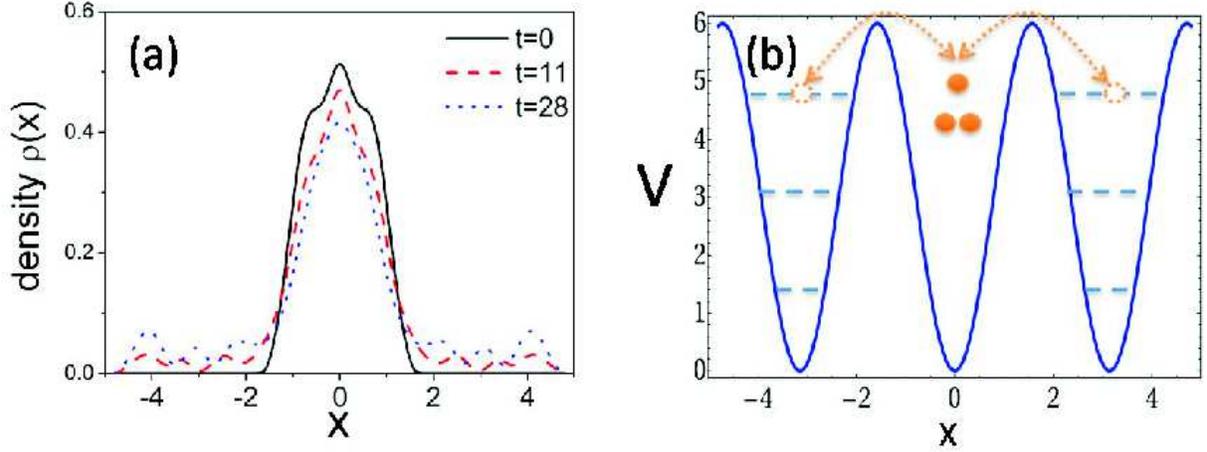}
\caption{(a) The one-body density at various time instants for the tunneling process in figure 5. All bosons are initially localized in the middle well (black line), then one boson tunnels to the left and right well with the same probability (red and blue lines). The nodal structure in the left and right well indicates the occupation of the second excited Wannier states in both wells . (b) Visualization of the tunneling process: in the tunneling process a boson tunnels from the middle well to the outer wells. The dashed lines in the left and right well illustrate the Wannier energy levels.
} \label{figure 6}
\end{figure}

In conclusion, we demonstrated that tunneling emerging from a state of  three bosons localized in the same well can be enhanced by the resonant coupling to number states relating to higher bands, and this resonant coupling gives rise to interband tunneling, in which a single boson tunnels to different excited Wannier states of different wells.

The above-mentioned interband tunneling can also interplay with the external confinement to achieve a tunable tunneling not only to a certain band but also to a certain well. A slight tilt potential $V_{tilt}=0.1\cdot x$ can \emph{e.g.} be applied to the triple well, which detunes $|2,1,0\rangle_{1}$, $|2,0,1\rangle_{1}$, and now $|3,0,0\rangle_{0}$ gets into resonance with $|2,1,0\rangle_{1}$ and $|2,0,1\rangle_{1}$ separately at different interaction strengths. Consequently a boson can selectively tunnel to the middle or right well at different interaction strengths. For instance, $|3,0,0\rangle_{0}$ gets into resonance with $|2,1,0\rangle_{1}$ at the particular interaction strength $g=3.0$, and consequently a single boson tunnels only to the middle well. Figure 7 provides the results of a tunneling process with tilt potential $V_{tilt}$, in which the boson only tunnels to the middle well at $g=3.0$.

 In experiments, the main difference of various realizations of the triple well potential is represented by the shift of the on-site energy of each well. The interband tunneling under tilt potential demonstrates that such tunneling is robust against the on-site energy shift, and consequently robust against deformations of the triple well potential from the ideal sine-square form. Earlier works also suggest that the tilt potential can be used to transport bosons in multiwell both dynamically \cite{Carr1} and adiabatically \cite{atomtronics}, and we point out that the interplay between the interaction potential and tilt potential can realize a specific transport of bosons not only into particular wells but also to particular Wannier states in corresponding wells.

 \begin{figure}
\includegraphics[width=16cm]{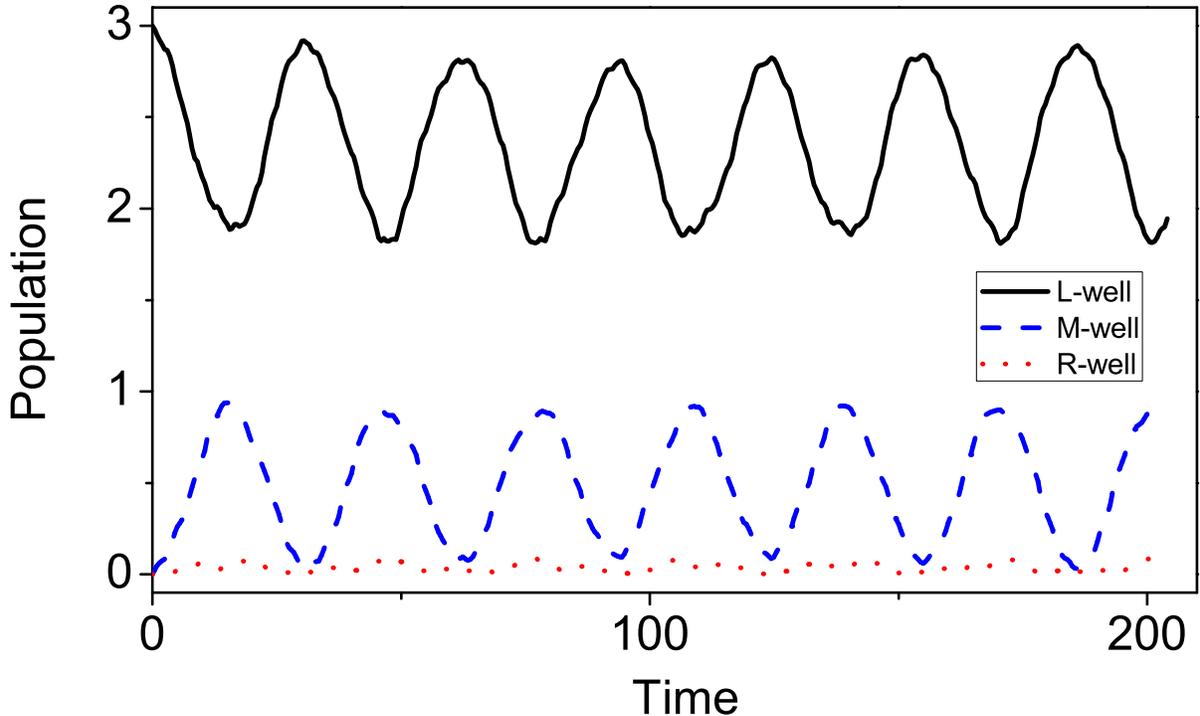}
\caption{Applying a tilt potential $V_{tilt}(x)=0.1\cdot x$, the tunneling of three bosons is shown with the initial condition that all bosons reside in the left well, for $g=3.0$. The population in the left, middle and right well is provided. One boson exclusively tunnels to the middle well.} \label{figure 7}
\end{figure}

\subsection{Two-boson correlated tunneling and beyond}
An interesting phenomenon in case of a double well is pair tunneling \cite{pair-exp,pair-tunneling-tho}, for which two repulsively interacting bosons tunnel as a pair forth and back. Pair tunneling arises since number states corresponding to bosonic pairs in each well are resonant with each other, and detuned  from all other states. Pair tunneling is a special case of correlated tunneling, in which particles do not tunnel independently. When higher bands come into play, various types of correlated tunneling occur and can be addressed by simply tuning the on-site energy of the initial state into resonance with the desired state.

\begin{figure}
\includegraphics[width=12cm]{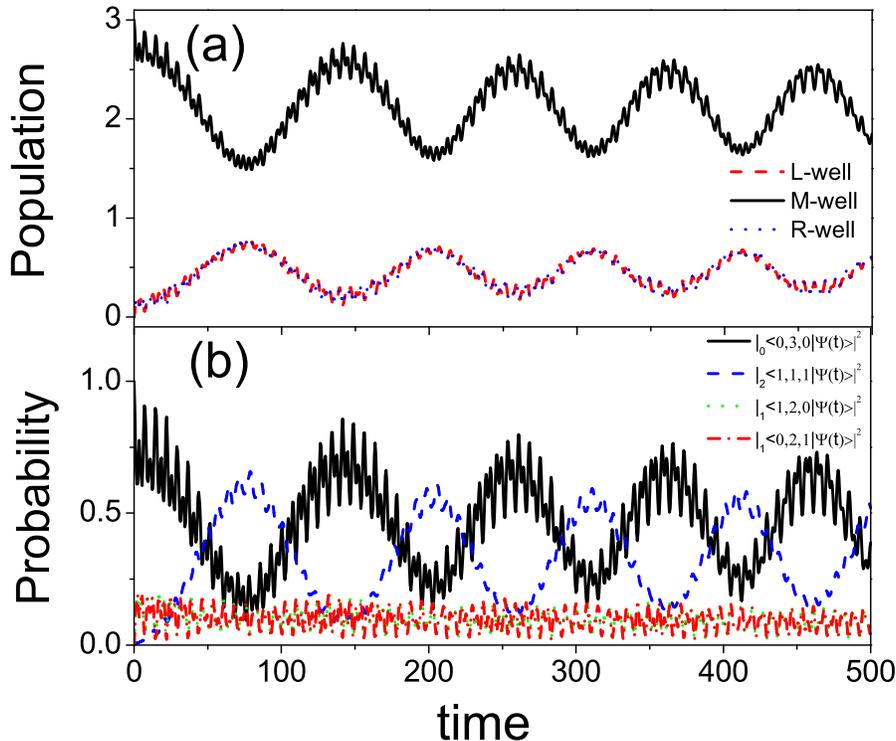}
\caption{Tunneling of three bosons initially localized in the middle well for $g=5.8$. (a) Population oscillation of the left well (red line), the middle well (black line), and the right well (blue line). The plots of the population in the left and right well practically lies on top of each other, indicating the correlated property of the two bosons tunneling to these wells. (b) Time occupation probability of the states $|0,3,0\rangle_{0}$, $|1,1,1\rangle_{6}$, $|1,2,0\rangle_{1}$ and $|0,2,1\rangle_{1}$. } \label{figure 8}
\end{figure}

 As an example, we demonstrate numerically a correlated tunneling process, in which two of the three bosons initially localized in the middle well simultaneously tunnel to the left and right well, respectively. We take as the initial state $|0,3,0\rangle_{0}$ for $g=5.8$, where $|0,3,0\rangle_{0}$ is resonant with $|1,1,1\rangle_{6}$, which refers to the configuration of a single boson located at the ground Wannier state of the middle well while the other two are in the first excited Wannier state of the left and right well, respectively. We show the evolution of the population in each well and the probability of number states in Figure 8. From the population oscillation shown in figure 8(a), we see that the occupation of the middle well oscillates within the interval $[1,3]$, and each occupation of the left and right well oscillates within $[0,1]$. This indicates that two bosons initially localized in the middle well propagate in a correlated manner to the left and right well in the course of the tunneling process. From the time evolution of the probability of number states, shown in figure 8(b), we find that the tunneling mainly takes place between $|0,3,0\rangle_{0}$ and $|1,1,1\rangle_{6}$, with a minor intermediate contribution from $|1,2,0\rangle_{1}$ and $|0,2,1\rangle_{1}$. According to the configuration $|1,1,1\rangle_{6}$, this means that during the tunneling two of the three bosons initially localized in the middle well tunnel to the first excited Wannier states of the left and right well simultaneously.

 In figure 9(a) we show the corresponding one-body density for different times to spatially resolve the interband tunneling process. The nodal pattern of the density profile in the left and right well confirms the occupation of the first excited Wannier states in these wells. Figure 9(b) is a schematic plot of such a tunneling process, based on the discussion above.

\begin{figure}
\includegraphics[width=16cm]{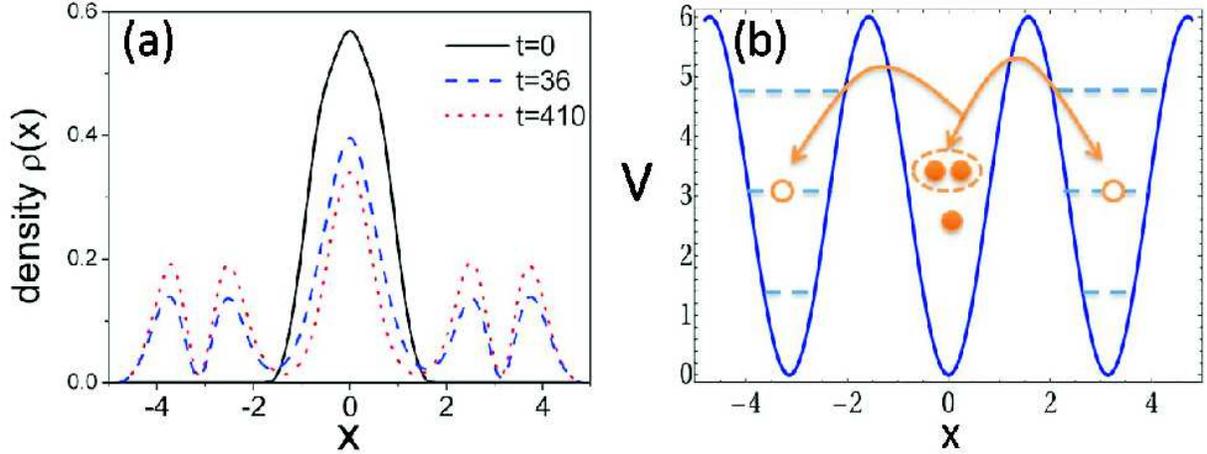}
\caption{(a) One-body density at different times for the tunneling as in Figure 8. Initially all the three bosons are localized in the middle well (black line), and gradually two of them tunnel to the left and right well (red and blue lines). (b) illustration of such tunneling process.} \label{figure 9}
\end{figure}

Correlated tunneling refers to the notion of tunneling of interacting particles that don't move independently, in comparison with the independent tunneling of free particles. In the single band approximation it is impossible to tune arbitrary number states into resonance via a change of the interaction strength, which is due to the different on-site energy dependence of number states on the interaction strength. This, however, becomes possible when we take into account higher bands, and in principle tunneling between arbitrary spatial and energetic configurations is achievable. As an example, we have shown here the correlated tunneling of two bosons initially in the middle well simultaneously to the left and right well.

Such interband tunneling can be straightforwardly extended to larger systems containing more bosons. We take four bosons in the triple well for instance. In such a system, there are four different categories of number states. In addition to the single mode, pair mode and triple mode, there exists the "quadruple modes" of $\{|4,0,0\rangle_{i},|0,4,0\rangle_{i},|0,0,4\rangle_{i}\}$, referring to four bosons in the same well. We can further separate the pair mode into two categories:  the "double-pair modes", such as $|2,2,0\rangle_{i}$, referring to the case that the four bosons are divided into two pairs and each of the pairs occupies a different well, and the "single-pair mode", containing a pair and two separate bosons, such as $|1,2,1\rangle_{i}$. Now in this system, besides the single-boson tunneling and two-boson correlated tunneling in the three-boson case, we can even realize correlated tunneling of pairs, by tuning the number states $|0,4,0\rangle_{0}$ and $|2,0,2\rangle_{1}$ into resonance, for instance.

\section{Summary and Conclusions}

We demonstrate that in a system consisting of a small ensemble of bosons confined in a one-dimensional triple well, several windows of enhanced tunneling are opened in the strong interaction regime, where in general the background of suppressed tunneling dominates. Such enhanced tunneling results from resonant coupling between the initial state and various excited number states, and can be only understood within an exact treatment of the system since it involves many bands, which is why we name it interband tunneling. When the initial state resonantly couples to different excited number states, various tunneling processes can be realized, which manifests itself as a potential tool for controllable dynamical transport of bosons. As an example, we demonstrate the single-boson tunneling to the first and second excited Wannier states, and the two-boson correlated tunneling in the system of three repulsively interacting bosons in the triple well with numerically exact quantum dynamical studies. The interband tunneling we discuss here can be straightforwardly generalized to multiwell systems with more bosons. In this way, tunneling between arbitrary spatial and energetic configurations of bosons in a  multiwell trap are in principle achievable just by tuning the interaction strength, and as a consequence controllable dynamical transport of bosons to specific wells and specific in-well energy levels becomes possible. This also opens the doorway to an interpretation and controllable preparation of the general non-equilibrium quantum dynamics in multiwell systems.

During the interband tunneling process, bosons tunnel to excited number states, and this gives the opportunity to couple the bosons to photons via induced emission processes. Such coupling can map the properties of number states, such as the spatial configuration, to emitted photons, and may manifest itself as a possible controllable photon source.

To understand the interband tunneling, we introduce a basis of generalized number states, which treat the interaction in a non-perturbative way such that these are valid even in the strong-interaction regime. Our number states are only meaningful and defined for large $V_{0}$. A main drawback of our method is that the interacting number states cannot be analytically generated, and we have to rely on a numerical approach to obtain them. However this cannot be considered as a serious drawback, since the gain in terms of analysis and transparent interpretation of the dynamical process is substantial, allowing us to understand and consequently design the non-equilibrium tunneling dynamics in optical lattices. We also notice that efforts have been put forward to provide an analytical description in the strong interaction regime \cite{multilevel,wannier}.

\begin{acknowledgments}
L.C. gratefully thanks the Alexander von Humboldt Foundation (Germany) for a grant. P.S. acknowledges the financial support from the Deutsche Forschungsgemeinschaft (DFG). Financial support from the German Academy of Science Leopoldina (grant LPDS 2009-11) is gratefully acknowledged by S.Z.
\end{acknowledgments}

\bibliographystyle{apsrev4-1}
\bibliography{reference}

\end{document}